\documentclass[journal=nalefd,manuscript=letter]{achemso}

\usepackage{amsmath,amssymb,amsfonts,amsthm}
\usepackage{graphicx}	

\def \ve{\varepsilon}
\def \pr{\parallel}

\author{Domitille Baux}
\affiliation[L2C]{Laboratoire Charles Coulomb, UMR5521, CNRS-Université de Montpellier, 34095 Montpellier, France}
\author{Patrick Hermet}
\affiliation[ICGM]{ICGM, Univ. Montpellier, CNRS, ENSCM, 34293 Montpellier, France}
\author{Stéphane Campidelli}
\affiliation[CEA]{Université Paris-Saclay, CEA, CNRS, NIMBE, LICSEN, 91191, Gif-sur-Yvette, France}
\author{Jean-Louis Bantignies}
\author{Emmanuel Rousseau}
\email{emmanuel.rousseau@umontpellier.fr}
\author{Nicolas Izard}
\email{nicolas.izard@umontpellier.fr}
\affiliation[L2C]{Laboratoire Charles Coulomb, UMR5521, CNRS-Université de Montpellier, 34095 Montpellier, France}

\title{Insights into the need for \textit{ab-initio} calculations to accurately predict the optical properties of metallic carbon nanotubes based on experimental confrontation}

\begin{document}

\begin{abstract}
In this article, we conduct comparative studies on the optical properties of metallic carbon nanotubes.
Firstly, we compare the complex dielectric constant predicted by an analytical model, the Linear Surface Conductivity Model, with \textit{ab initio} calculations based on Density Functional Theory.
We highlight the similarities and differences between these two models, with the major discrepancy being a significantly different behavior of the plasma frequency with respect to the carbon nanotube diameter.
In the second step, we compare the predictions of these models with experimental measurements of the dielectric function.
We demonstrate that the screened plasma frequency serves as a reliable quantifier for distinguishing between the two models.
In conclusion, we find that the \textit{ab initio} calculations more accurately describe the optical properties of metallic carbon nanotubes compared to the commonly used Linear Surface Conductivity Model.
\end{abstract}

\section{Introduction}
Carbon nanotubes are appealing nanostructures made exclusively of carbon atoms that display a large variety of electronic and optical responses.
Depending on their structure, carbon nanotubes could display metallic or semiconducting electronic properties.
Semiconducting carbon nanotubes have been extensively studied because of their potential as light sources in the near-infrared spectral range\cite{Avouris_2008, Gaufres_2012}.
On the other hand, metallic Single-Wall Carbon NanoTubes (m-SWCNT) could be envisioned as the smallest metallic wire at the nanoscale with exceptional electrical properties.
Indeed, in contrast to nanowires based on noble metals\cite{Pearson_2012}, individual m-SWCNTs are defect-free and present the highest-known DC-conductance\cite{Poncharal_2002}.
The potential of applications of carbon nanotubes is not restricted to optic or electronic areas\cite{Avouris_2008} but also includes biological and medical studies\cite{Prato_2007} where their small size allows for drug delivery inside cells.
Quantitative predictions of many physical phenomena implying carbon nanotubes requires the knowledge of their optical properties.
Some typical examples are the potential use of a single m-SWCNT as a plasmonic antenna\cite{Lakhtakia_2006} or playing the role of an optical cavity to modify the spontaneous emission rate of an emitter due to Purcell effect\cite{Bondarev_2002, Lakhtakia_2010}.
Some applications need the knowledge of SWCNT optical properties on a wide frequency range, the radiative heat transfer between two carbon nanotubes\cite{Nemilentsau_2007,Rubi_2022} being an example. A second one are Van der Waals forces that dominates interactions at the molecular scale. Being of electromagnetic nature, their magnitude as well as their possibility to be repulsive or attractive crucially depend on the optical properties of the molecules and their environment\cite{Hutter_1993}.
A quantitative calculation of Van der Waals forces acting on carbon nanotubes\cite{Rubi_2021} may be valuable giving new insights for any application in biology.
Both the real part and the imaginary part of the dielectric function are required to make meaningful predictions of the above applications.

In this article, we focus on the optical properties of m-SWCNT because they have been little explored compared to those of semiconductor carbon nanotubes\cite{Avouris_2008}.
Specifically, we want to unravel the contribution of the intraband transitions that lead to the metallic behaviour of m-SWCNT which is of primary interest in several applications such as plasmonic antennas\cite{Lakhtakia_2006}, radiative heat transfer\cite{Nemilentsau_2007,Rubi_2021} or biology\cite{Rubi_2022}.

\section{Theory and Experiment}
The optical properties of materials can be determined through experimental measurements or by modelling their response to electromagnetic field excitations.
However, it is crucial to validate models by comparing them with the experimental data in order to assess their accuracy and the range of their validity.
Regarding m-SWCNT, the literature reports both theoretical studies and experimental works.
Theoretical investigations encompass a range of approaches, including first principles numerical calculations\cite{Duan_2004,Halim_2021} as well as  analytical methods\cite{Slepyan_1999,Hanson_2005}.
Most of the \textit{ab initio} studies focused on the optical properties of semiconducting SWCNT to predict their band-gap\cite{Zolyomi_2004,Gharbavi_2016,Rai_2020}.
While there is some works concerning m-SWCNTs, a crucial aspect, namely the contribution of intraband transitions that accounts for the metallic behavior of m-SWCNTs, is often missing\cite{Gharbavi_2015}. T. Movlarooy\cite{Movlarooy_2013} computed the dielectric functions for two metallic chiralities, (8,8) and (15,0); however, the trend of variations in optical properties as a function of diameter was not provided, making it difficult to extrapolate their behavior for other chiralities.
The linear surface conductivity model is a widely used analytical model\cite{Slepyan_1999} for describing the optical properties of carbon nanotubes.
This model is applicable to both semiconducting and metallic chiralities, but it relies on approximations and has not been validated through comparison with measurements over a wide frequency range. It is important to emphasize that in order to conduct a meaningful comparison between model's predictions and experimental measurements, the following requirements need to be met:
\begin{itemize}
\item The sample must be sorted by separating the metallic and the semiconducting chiralities\cite{Tanaka_2011}. This assures that the measurement of the dielectric function of metallic chiralities is not spoiled by the contribution of the semiconducting chiralities and vice versa.
\item The real part and imaginary part of the dielectric constant have to be measured as they both contribute to the electrodynamics properties of m-SWCNT heterostructures\cite{Lakhtakia_2006,Lakhtakia_2010,Rubi_2022}. Measuring a single quantity, such as the scattering cross section\cite{Sfeir_2006}, is not enough to be conclusive as errors in the real part and imaginary part may compensate to reproduce the measurement.
\item In the infrared range and at smaller frequencies, the dielectric constant of m-SWCNT follows a Drude's model.  Its variations typically scale as $1/\omega$, where $\omega$ is the light frequency. A measurement that aims to be quantitative have to be performed over several frequency decades.  
\end{itemize}

A review of the literature shows that none of the available measurements fulfills these requirements. Indeed some measurements of the dielectric function are limited to unsorted samples mixing semiconducting and metallic SWCNTs\cite{Ugawa_1999, Zhou_2005, Borondics_2006, Pekker_2011, Maine_2012, Ermolaev_2020, Song_2020}. Ref.\cite{Tohati_2014} reports the real-part of the optical conductivity for m-SWCNT over a wide frequency range but missed the imaginary part.
The most complete report to date\cite{Zhang_2013} has measured the complex electrical-conductivity for metallic and semiconducting SWCNTs, but only for a narrow frequency range (0-40~THz, 0-0.16~eV). 

The goal of this paper is two-fold. Firstly, we wish to quantitatively compare the predictions from the linear surface conductivity model and from \textit{ab initio} calculations since this last theoretical method can relax some assumptions performed to tackle the analytical model.
We report new theoretical values for the  plasma frequency of m-SWCNT.
First-principles based calculations predict a linear dependency of the plasma frequency with the m-SWCNT diameter while the linear surface conductivity model predicts variations as the inverse of the m-SWCNT diameter.
Secondly, we wish to fill the lack of comparison between theoretical predictions and measurements.
We have measured both the real part and the imaginary part of the dielectric function over the energy range [0.05,4] eV for a m-SWCNT sample. We consider the screened plasma frequency $E^\star$ as a quantifier to unambiguously prove that the first principles calculations predict more accurately the optical properties of m-SWCNT than the linear surface conductivity model.

Carbon nanotubes are anisotropic materials, and their dielectric function $\bar \ve(\omega)$ at the angular frequency $\omega$ is a second rank tensor with two different contributions: 

\begin{align*}
\bar \ve(\omega) = 
\begin{pmatrix}
 \ve^\bot(\omega) & 0 & 0\\
 0 &   \ve^\bot(\omega) & 0\\
 0 & 0 &  \ve^\pr(\omega)  
 \end{pmatrix}
\end{align*}

where $\ve^\pr$ is the axial dielectric function, along the nanotube axis and $\ve^\bot$ is the transverse dielectric function in the plane perpendicular to the SWCNT axis.
Both intraband and interband transitions\cite{Burdanova_2021} contribute to the dielectric constant.
Generally, we can write each component of the dielectric tensor as being the sum of the contribution of intraband and interband transitions: $\ve^\pr(\omega) = \ve^\pr_{intra}(\omega) + \ve^\pr_{inter}(\omega)$  and $\ve^\bot(\omega) = \ve^\bot_{intra}(\omega) + \ve^\bot_{inter}(\omega)$.
These last four quantities can be independently computed either by the linear surface conductivity model or by \textit{ab initio} methods.

The linear surface conductivity model is analytical. It assumes that carbon nanotubes are strictly one-dimensional objects being infinitely thin (Assumption A1) and infinitely longs.
The immediate consequence of the assumption (A1) is that carbon nanotubes are \textit{non-polarizable object} in the directions transverse to the cylinder axis. Consequently, the transverse dielectric constant is fixed to one, \textit{i.e.} $\ve^\bot(\omega)= 1$, in this model.

The linear surface conductivity model assumes that the electronic response to an electromagnetic wave excitation is dominated by the $\pi$-electrons, neglecting the contribution of $\sigma$-electron. The dispersion relations of the $\pi$-electrons are computed in the tight-binding approximation\cite{Saito} assuming the parameters to be those of graphene.
This approximation neglects the curvature effects inherent to the geometry of carbon nanotubes. These assumptions are referred as the set of approximations (A2).
These assumptions are performed to process the analytical model and it is difficult to associate them individually with the contribution of the intraband or the interband transitions to optical properties of CNT.
The assumption A1 implies that the carbon nanotubes are infinitely thin and infinitely long object, as a 1D wire. This has a direct consequence on the perpendicular dielectric function for both the intraband and the interband transitions. It implies that the transverse dielectric function is strictly equal to one.
On the other hand, the assumption A2 involves calculating the electronic dispersion relation in the tight-binding approximation with the graphene parameters. Consequently, only the $\pi$-electrons contribute to the optical properties while the contribution of the $\sigma$-electrons is neglected. This assumption also neglects the curvature effects that are known to be non negligible at low energy due to a curvature-induced gap for semi-metallic chiralities. This assumption has consequences on both the intraband and the interband transitions.

The amplitude of the external electromagnetic field is assumed to be small leading to linear perturbation of the electronic distribution function around the electronic distribution function at equilibrium. So, under this approximation, both the contribution from the intraband and the interband transitions can be analytically computed from the Boltzmann equation. They are given by the following equations:

\begin{align}
	\sigma_1(\omega) = - \frac{i e^2}{2 \pi^2 R_{CNT} \hbar (\omega +i/\tau_1)} \sum_{k=1}^{m} \int_{-a}^{a} \frac{\partial \ve_c(p_z,k)}{\partial p_z} \frac{\partial \rho^{eq}(p_z,k)}{\partial p_z} d p_z
\end{align}

\begin{align}
	\sigma_2(\omega) = \frac{i e^2}{\pi^2 R_{CNT} \hbar} \sum_{k=1}^{m} \int_{-a}^{a} \frac{\omega_{cv}(p_z,k) P_{cv}^2(p_z,k) \rho^{eq}(p_z,k)(\omega + i / \tau_2)}{\omega^2_{cv}(p_z,k) - (\omega + i/\tau_2)^2} d p_z
\end{align}

respectively for the intraband and the interband transitions\cite{Nemilentsau_2011}.
The quantity $\ve_c(p_z,k)$ represents the dispersion law of $\pi$-electrons in the SWNT where $a$ defines the first Brillouin zone.
The integration is performed over the variable $p_z$, which denotes the projection of the electron quasimomentum onto the SWNT axis.
The quantities $P_{cv}$ and $\omega_{cv}$ refer to the normalized matrix element of the electron dipole momentum operator and the frequency of the electron interband transitions, respectively. $\rho^{eq}$ corresponds to the population difference between the conduction band and the valence band at thermal equilibrium. $R_{CNT}$ denotes the carbon nanotube radius, while $\tau_1$ and $\tau_2$ are the relaxation time for the intraband and interband transition, respectively. Finally, $e$ is the electron charge and $\hbar$ is the reduced Planck constant.

The linear surface conductivity $\sigma^\pr$ has the dimension of a conductance. Consequently, a characteristic length-scale $t^\star $ has to be introduced to relate the surface conductivity $\sigma^\pr$ and the dielectric function $\ve^\pr$. It is an \textit{ad-hoc} parameter. We follow ref.\cite{Slepyan_2010} and choose the carbon nanotube diameter $t^\star = 2R_{CNT}$ as this ad-hoc parameter. Consequently, the dielectric function depends on the linear surface conductivity $\sigma^\pr(\omega)$ through the equation:
\begin{align}
	\ve^\parallel(\omega) = 1+i \frac{\sigma^\pr(\omega)}{ \ve_0 t^\star \omega}
	\label{eq:epsB}
\end{align}
where $\ve_0$ is the vacuum permittivity.

The linear surface conductivity model can deal with semiconductors and metallic carbon nanotubes but the computation of the interband transitions is practically limited to achiral SWCNT. Indeed, only for achiral nanotubes, the energy-dispersion relations for the valence and the conduction bands are explicitly known.\cite{Saito}. At low energy, the contribution of the intraband transition dominates the contribution of the interband transitions. This contribution is given by a Drude's model\cite{Kittel}, a generic model describing the optical properties of metals. Drude's model depends on two parameters, the plasma frequency $\omega_p$ and the relaxation time $\tau$. The linear surface conductivity model allows the computation of the plasma frequency through the equation:
\begin{align}
	\omega_p = \sqrt{\frac{3 b \gamma_0 e^2}{\pi^2 \hbar^2 \ve_0 t^\star} \frac{1}{R_{CNT} }}
	\label{eq:freqPlas}
\end{align}

Where $b=0.142$ nm is the interatomic distance between carbon atoms, $\gamma_0=2.7$ eV is the transfer integral and $R_{CNT} = \frac{b\sqrt{3}}{2\pi}\sqrt{m^2+n^2+m n}$ is the radius of a SWCNT with chiral indices (m,n).
Eq.(\ref{eq:freqPlas}) is valid for achiral m-SWCNTs\cite{Nemilentsau_2011} but also for chiral m-SWCNTs\cite{Slepyan_1999}.

The relaxation effects are incorporated phenomenologically\cite[p.154]{Maksimenko_2004}.
Although a different relaxation time can be introduced for the intraband $\tau_1$ and the interband transitions $\tau_2$, they are usually assumed to be equals\cite{Maksimenko_2004, Slepyan_2010, Nemilentsau_2011, Shuba_2012, Rubi_2022}, $\tau_1=\tau_2$. To simplify notations, they will be denoted by $\tau$ in the following. We will assume $\tau= 35 \text{~fs}$ in agreement with ref.\cite{Maksimenko_2004,Nemilentsau_2011}.
Note that this value is consistent with our experimental measurements as explained in the following.

In this paper, we have computed the optical properties of m-SWCNTs taking advantage of the Density Functional Theory (DFT). Kohn-Sham equations are numerically solved using a self consistent procedure. 
This method relaxes the assumptions (A1) and (A2) inherent to the analytical model: SWCNTs are no more assumed to be 1D-object.
The electronic potential takes into account both the $\sigma$-electrons and the $\pi$-electrons. It is not limited to the first three adjacent atoms in the hexagonal structure.
Furthermore, DFT takes into account curvature effects by computing a selected chirality (n,m). 
DFT calculations predicts both the intraband and interband contributions to the dielectric function.
Since we focus the study on m-SWCNT, we are not interested on excitons contribution.
We show in Supplementary material S4 that the excitons contribute only marginally to the dielectric function as compared to the contribution of the interband transitions.
The optical spectra of the different nanotubes are obtained using the first-order time-dependent perturbation theory within the random phase approximation (RPA).

For the interband contributions, the diagonal terms of the imaginary part of the dielectric function, $Im[\varepsilon_{inter}^\sigma]$ with $\sigma=\parallel$ or $\bot$, can be calculated from the eigenvalues and wavefunctions of a band calculation: 
\begin{equation}
	Im[\varepsilon_{inter}^\sigma(\omega)]=\left(\frac{e}{\pi m\omega} \right)^2\sum_{i,f}\int d^3k\  |<i\mathbf{k}|P^\sigma|f\mathbf{k}>|^2\  f_{i\mathbf{k}}(1-f_{f\mathbf{k}})\  \delta(E_f(\mathbf{k})-E_i(\mathbf{k})-\hbar\omega),
\end{equation}
where $i$ and $f$ respectively indicate the initial and final states, $|i\mathbf{k}>$ is the eigenstate with wavevector $\mathbf{k}$ and band index $i$, $E_i(\mathbf{k})$ is the corresponding eigenvalue, $f_{i\mathbf{k}}$ denotes the Fermi distribution, $P^\sigma$ is the $\sigma$-component of the momentum operator, $\hbar\omega$ is the photon energy, $m$ is the electron mass and $\delta$ is the Dirac distribution. The real part of the dielectric function was obtained from Kramers-Kronig relations. 

The contribution of the intraband transitions is treated by a Drude model. The plasma frequency is calculated according to:
\begin{equation}
	\omega_p^2=\frac{e^2}{\pi^2\hbar^2 } \sum_i\int d^3k\ \left(\frac{\partial E_i(\mathbf{k})}{\partial\mathbf{k}}\right)^2\ \delta(E_i(\mathbf{k})-E_F)
\end{equation} 
where $E_F$ is the Fermi energy.

Similarly to the linear surface conductivity model, DFT calculations do not predict relaxation effects at the RPA level.
Consequently, the DFT calculations may overestimate the contribution of the interband transitions.
Concerning the intraband transitions, they are incorporated phenomenologically through relaxation time $\tau$. The numerical value of $\tau=35$ fs is used as for the linear surface conductivity model.

More details on the computation of the dielectric function with the DFT method are given in supplementary material.

\begin{figure}
	\includegraphics[width=\linewidth, clip]{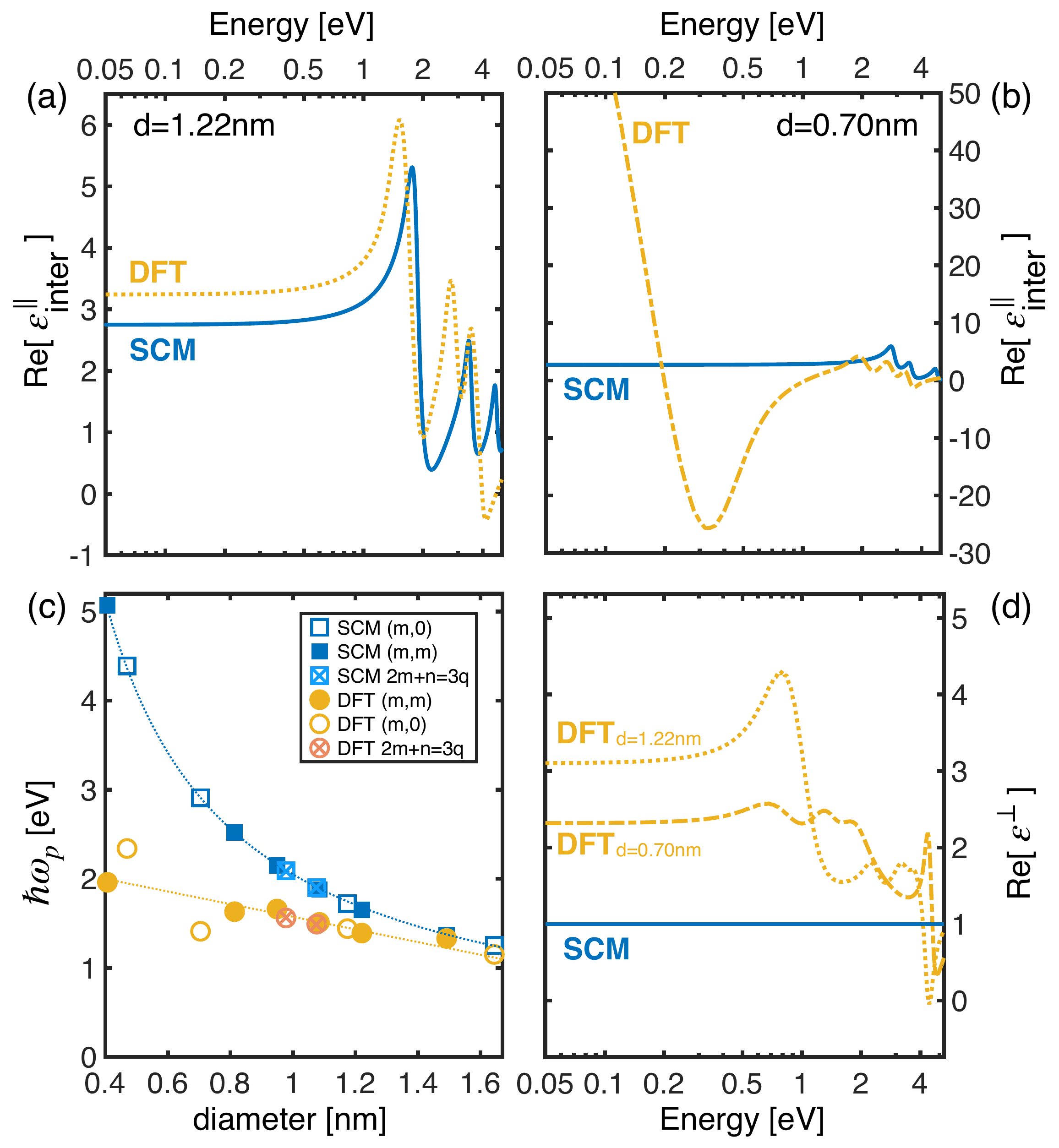}
	\caption{Contribution of the interband transitions to the real part of the dielectric constant as calculated by the DFT method (yellow dashed line) or the linear surface conductivity model (SCM, solid blue line)  (a) for d=1.22 nm [chirality (9,9)] (b) d=0.70 nm [chirality (9,0)] (c) Plasma frequency as a function of the m-SWCNT diameters predicted from the DFT calculations (filled circles) and  the surface conductivity model (filled squares). Results from DFT are fitted with a linear curve and results from the linear surface conductivity model follow a hyperbola. (d) Real-part of the interband contribution of the transverse component of the dielectric tensor. Dashed lines are predictions from DFT while the solid line is the assumption from the linear surface conductivity model.} 
	\label{fig1} 
\end{figure}

Results from the two theories are compared in Fig.(\ref{fig1}).
First we discuss the axial component of the dielectric tensor $\ve^\pr(\omega)$ for both the interband and the intraband transitions.
Fig.(\ref{fig1}-a) shows the real part of the contribution of the interband transitions to the dielectric function for a m-SWCNT with diameter $d=1.22$ nm [chiral indices (9,9)].
The curves are similar with a slight difference at low energy but with a 0.4~eV shift at the first peak. Higher order peaks present a larger discrepancy.
Fig.(\ref{fig1}-b) shows the real part of the contribution of the interband transitions to the dielectric function for a m-SWCNT with diameter $d=0.70$ nm [chiral indices (9,0)].
This chirality is of semi-metallic nature, characterized by a band-gap at low energy, measured in ref.\cite{Ouyang_2001} near $E \sim 0.08$ eV in agreement with our DFT calculations.
This band-gap manifests itself as a large contribution in the real and imaginary part of the dielectric constant at low energy and is responsible for the low-energy blow-up observed in Fig.(\ref{fig1}b).
The linear surface conductivity model does not predict this contribution since the low-energy band-gap results from the m-SWCNT curvature.
This discrepancy between the two theories is not negligible. 

The intraband transitions contribute to the dielectric function as a Drude model in the two theories. As a consequence, we compare the predictions for the plasma frequency $\omega_p$.
They are shown in Fig.(\ref{fig1}-c) as a function of the m-SWCNT diameters.
The calculations has been done for metallic nanotubes with chiral indices $(m,m)$ and for semi-metallic nanotubes with chiral indices $(m,0)$ or $(m,n)$ such as $2m+n=3q$ with $q \in \mathbb{N}$.
The predictions from DFT scale linearly with carbon nanotubes diameter while the predictions from the linear surface conductivity model scale as $\propto 1/d$ in assuming $t^\star = 2 R_{CNT}$ in Eq.(\ref{eq:freqPlas}).
While the two curves seem to converge to similar values as the diameter goes to 1.6 nm, the deviation in the predictions is significant for diameters smaller than 1.2 nm. Again, the two theoretical methods predict rather different values for the intraband contributions.

For the transverse component $\ve^\bot$ of the dielectric tensor, the two theories predict that there is no contribution of the intraband transition, the plasma frequency being null.
The contribution of the interband transitions is compared in Fig.(\ref{fig1}-d).
The yellow dashed curves are the predictions from the DFT calculations for m-SWCNT with diameter 1.22 nm and with diameter 0.7 nm.
The prediction from the linear surface conductivity model is shown as the solid blue curve. While this model predicts a contribution equals to $\ve^\bot(\omega) = 1$ independently of the diameter as a result of the approximation (A1), the DFT calculations predict radically different results as it can be observed in Fig.(\ref{fig1}-d).
They are characterized by a plateau below the near-infrared range and displays absorption bands at higher energies.

All our results suggest that relaxing the approximation (A1) and (A2) leads to different predictions, some being significantly different such as the plasma frequency or the transverse component $\ve^\bot$ of the dielectric tensor.
We will now compare these theoretical predictions with experimental measurements.

\begin{figure}
	\includegraphics[width=\linewidth]{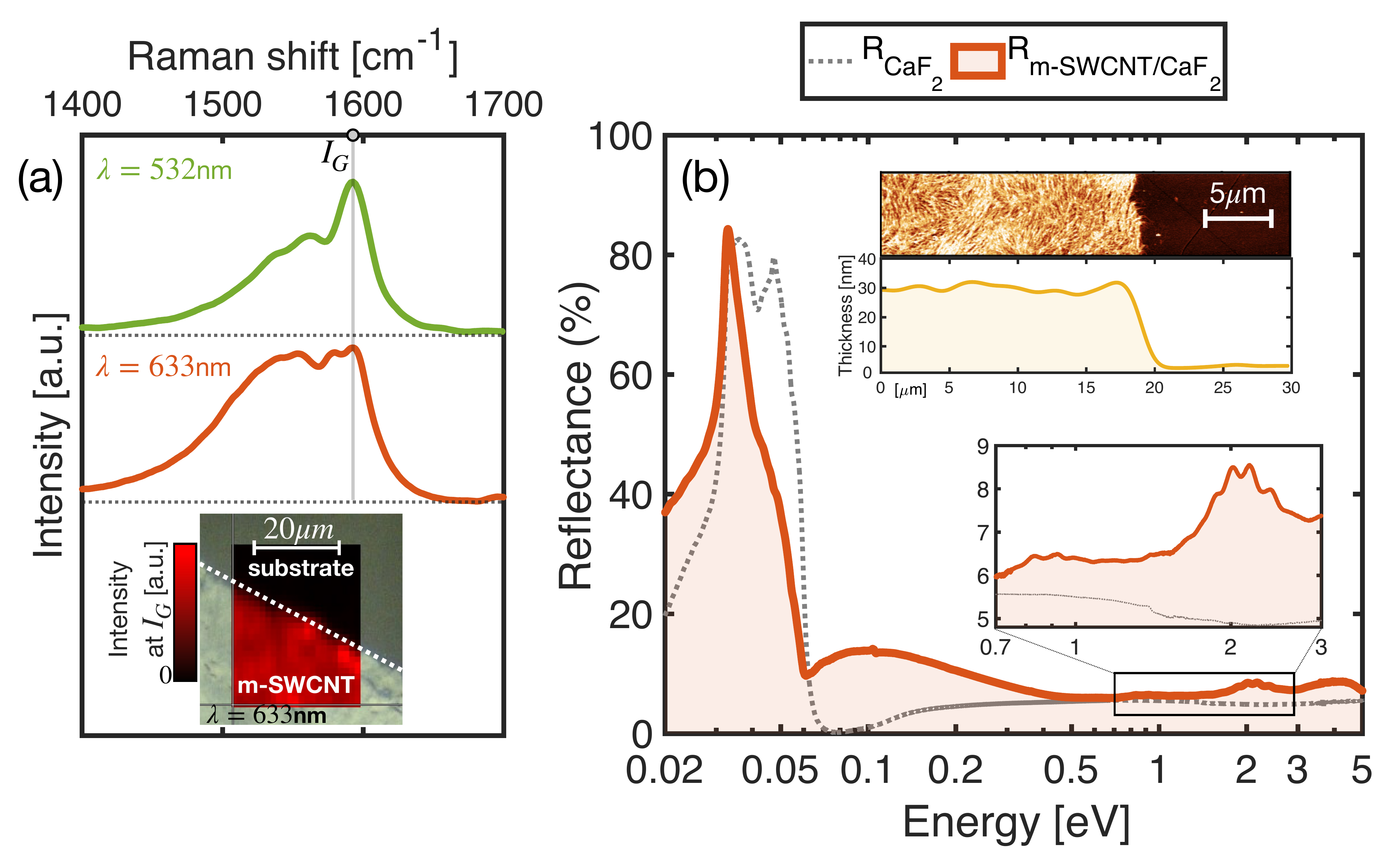}
	\caption{(a) Typical Raman spectra for excitation at 633~nm (red curve) and 532~nm (green curve). The inset shows a mapping of the G band at 1590 cm$^{-1}$ over the m-SWCNT film (excitation at 633nm). (b) Experimental reflectance of a m-SWCNT film on CaF$_2$ substrate (red curve) and of the reference CaF$_2$ substrate (dot curve). The high energy range is zoomed to observe the excitonic transitions. Inset upper panel: AFM image of a typical m-SWCNT film. Inset lower panel: mean thickness obtained by averaging on a 5~$\mu$m window.}
	\label{fig2}
\end{figure}

The m-SWCNT sample is based on HiPCO carbon nanotubes (Nano Integris). They are sorted by column chromatography following the procedure described by Tanaka et al. \cite{Tanaka_2011} to obtain enriched m-SWCNT suspensions.
They are filtered onto nitrocellulose membranes to form thin films which are then transferred onto calcium fluoride $\text{CaF}_2$ substrates.
Samples are further annealed under high vacuum ($10^{-6}$~Torr) at 250$^{o}$C for 4 hours to remove solvent and impurities\cite{Pekker_2011,Zhang_2013}.
The film's thickness, $d_1$, is determined by Atomic Force Microscopy (AFM) to be $30 \pm 5 \text{~nm}$ (inset in Fig.\ref{fig2}(b)).
Raman spectra (Fig.\ref{fig2}(a)) performed at 532 and 633~nm feature a strong peak at $1590 \text{~cm}^{-1}$, the so-called G-band, typical of SWCNTs.
The broad asymmetric Fano lineshape below the G-band is a key characteristic of metallic carbon nanotubes\cite{Hasdeo_2013}, assessing the sorting quality.
The inset in Fig.\ref{fig2}(a) displays a mapping of the intensity of the G-band, underlying in conjunction with the AFM data the nanotube film homogeneity in terms of thickness, density, and nanotube orientation.
It is important to underline that no m-SWCNT preferential orientation inside the layer was evidenced.
The m-SWCNT diameter distribution is relatively broad but could be bracketed between 0.7 and 1.2~nm\cite{Chiang_2001} (See supplementary material S1).

Broadband reflectance spectra are measured using a FTIR spectrometer (Bruker IFS 66v/S) adapted with a combined transmission and specular reflection apparatus from 0.02~eV up to 0.6~eV, and a UV-Vis-NIR spectrometer (Cary Varian 5000) adapted with a VW absolute specular reflectance accessory from 0.6 eV up to 5 eV. Measurements are done at near-normal incidence in both cases.
Great care is taken to acquire the reflectance spectra which requires an extremely fine alignment procedure to achieve a precision on the order of 1~\%.
Measurements are calibrated on materials whose optical properties are known.
Fig.\ref{fig2}(b) displays the m-SWCNT film reflectance deposited on a CaF$_2$ substrate (red curve), while the bare substrate reflectance is the dotted-gray curve.
The 0.7 to 3~eV range is magnified to display the carbon nanotube excitonic transitions contribution\cite{Miyata_2008,Pekker_2011}.
While the M$_{11}$ transitions (near 2 eV) contribute to the reflectance largely above the level due to the substrate contribution, the S$_{11}$ transitions (0.7-1.2~eV) contribute marginally to this reflectance, underlining again sorting efficiency (see supplementary material S2 for additional comparison of the absorption spectra before and after sorting).

The complex dielectric constant of the m-SWCNT layer is extracted from the reflectance measurements using Kramers-Kronig relations\cite{Jahoda_1957,Nash_1995,Rousseau_2021} relating the phase $\theta$ of a complex number $z=|z|e^{i\theta}$ to its modulus $|z|$. 
The procedure requires the determination of the complex Fresnel coefficient $r_{02}$ which describes light reflection by the m-SWCNT layer deposited on a semi-infinite CaF$_2$ substrate.
The reflectance experiment does not measure $|r_{02}|^2$ directly. While thick compared to any wavelength probes in this letter, the finite size of the substrate ($d_2 = 1$ mm) has to be taken into account, particularly in the spectral range where CaF$_2$ is transparent.
Supplementary material S3 explains how to relate the measured reflectance $R$ to the quantity $|r_{02}|^2$.

Once the modulus $|r_{02}|$ and the phase $\theta_{02}$ of $r_{02}$ are known, the m-SWCNT refractive index could be extracted by solving:
\begin{equation}
\label{eq1}
r_{02}[\tilde n_1(\omega)] = |r_{02}| e^{i \theta_{02}(\omega)}
\end{equation}
where $r_{02}[\tilde n_1(\omega)]$ is the analytical expression of the Fresnel coefficient. The unknown to be found is the m-SWCNT complex refractive-index, $\tilde n_1(\omega)$. The quantity on the right side of Eq.~(\ref{eq1}) is derived from the experiment.

\section{Results and discussion}

\begin{figure}
	\includegraphics[width=\linewidth]{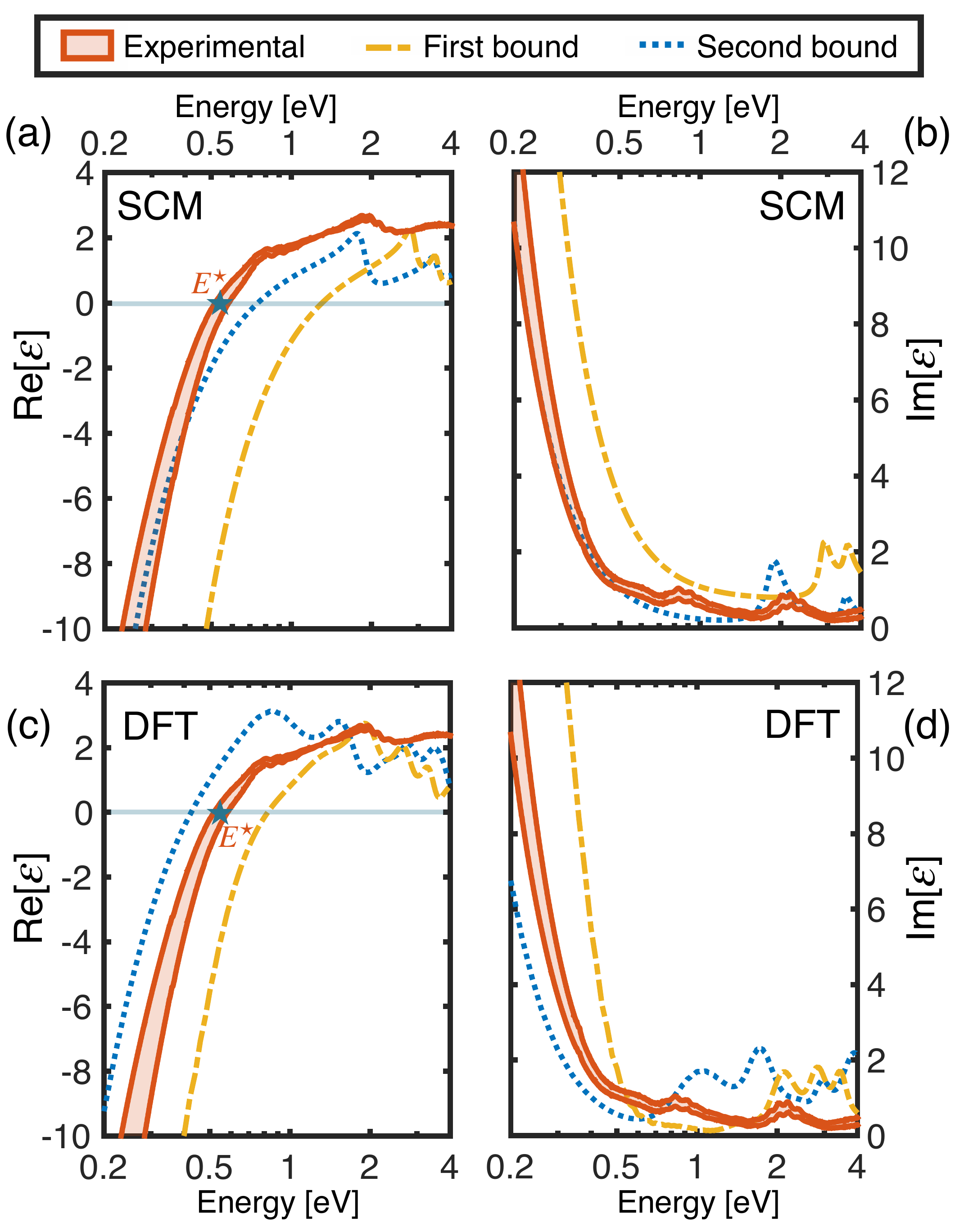}
	\caption{Experimental measurement of the (a,c) real part and (b,d) imaginary part of the dielectric constant of the m-SWCNT layer compared with predictions from (a)(b) the surface conductivity model (SCM) and (c)(d) DFT calculations. The surface conductivity model assumed the parameters: $\gamma_0=2.7$ eV, $\tau = 35$ fs and $T=300$ K. Experimental measurements are in light-red shaded area, including confidence interval. Theoretical predictions for the first and second bounds are represented as dotted lines. All curves cross the zero-line (grey line) at the cut-off energy $E^\star$ shown as the purple star for the experiment.}
	\label{fig3}
\end{figure}

The experimental dielectric constant found from this procedure is plotted in Fig.\ref{fig3}, real and imaginary parts respectively on the left and right side. The experimental confidence interval is represented by a shaded area. This area includes the uncertainties on the film thickness, the substrate refractive index, the reflectance spectral join between the NIR / MIR range and the evaluation of $\theta$, the phase of the Fresnel coefficient. The main source of errors comes from the uncertainty on the film thickness.
To check the accuracy of the experimentally determined dielectric constant, the sample transmittance was measured in the 0.05 to 4~eV range and compared with predictions based on this experimental dielectric constant as shown in section S4 in supplementary materials.
This measurement, independent from the one used to extract the dielectric constant, shows a good agreement with the predictions. It is a blind test to check the consistency of the dielectric constant derived from the reflectance measurements.

We now aim at comparing the experimental optical properties of m-SWCNTs with theoretical predictions. Carbon nanotubes form a composite with many chiralities of metallic nature. The composite has the geometry of a flat layer deposited on a substrate. Inside the layer, carbon nanotubes have a complicated geometry. They are not rigid fibres being neither perfectly straight nor aligned as shown in the AFM picture [Fig.(\ref{fig2}-b)]. The geometry of carbon nanotubes' layers has been qualified as "tumbleweed configurations"\cite{Ma_2008}. Consequently, at the microscopic scale, the layer can be understood as a succession of polycrystalline clusters without correlations at the macroscopic scale. The layer can be seen as a powder made of crystallites with random orientation. The crystallites' characteristic size is roughly on the order of one to a few tens of nanometers. They are then extremely small compared to the wavelength and the size of the incident light beam. Consequently, the layer behaves as a homogeneous, isotropic material under light illumination. The effective dielectric constant of the layer is the arithmetic average\cite{Andersen_2002} of $\ve^\bot(\omega)$ and $\ve^\pr(\omega)$:
\begin{align}
\ve(\omega) = \frac{2}{3} \ve^\bot(\omega) + \frac{1}{3} \ve^\pr(\omega)
\label{eq:epsA}
\end{align}

Eq.(\ref{eq:epsA}) has been used to describe the optical properties of polycrystalline graphitic clusters\cite{Andersen_2002,Draine_1984}, layers of multiwall carbon nanotubes\cite{Wasik_2017} and layers of single-wall carbon nanotubes\cite{Slepyan_2010}.

In Fig.(\ref{fig3}), we compare the experimental data (red shaded area) with either the predictions from the linear surface conductivity model [Fig.\ref{fig3}(a) and (b)] or with the results from the DFT calculations [Fig.\ref{fig3}(c) and (d)]. The real part and the imaginary part are shown in Fig.\ref{fig3}(a-c) and Fig.\ref{fig3}(b-d), respectively. 

Since neither the surface conductivity model nor the DFT calculations predict the relaxation time $\tau$ for the intraband transitions, it is introduced as a phenomenological parameter in the two theories. We used the same numerical value of $\tau= 35 \text{~fs}$ in the two theories\cite{Slepyan_1999,Maksimenko_2004,Slepyan_2010}. This value is extracted from the experimental curves from a fit with a Drude model in the range [0.05,0.5] eV (See supplementary material S6 for more details). It is in excellent agreement with the value given in the literature\cite{Nemilentsau_2011}.  

While being sorted and enriched in metallic chiralities, the sample contains a distribution of metallic nanotube species, armchair ($n$,$n$), zigzag ($n$,0) and chiral ($n$,$m$), in the diameter range determined from the Radial Breathing Modes study (See supplementary material S1).
To compare the experimental data with predictions, on the basis of the trends of variations given by the two theories [see Fig(\ref{fig1})], we define two boundaries for the dielectric function.

The first bound is defined as:
\begin{itemize}
	\item A chirality with a diameter close to 0.70 nm, the minimal diameter in the diameter distribution.   
	\item A semi-metallic chirality since the gap at low energy shifts the dielectric function curve to higher values as compared to a metallic chirality without low-energy gap in the infrared.
\end{itemize}

Based on these arguments, we choose the chirality (9,0) as the first bound to compare with the experiment. The predictions for the first bound are shown as the dashed yellow line in Fig.(\ref{fig3}).

The second bound is defined as:
\begin{itemize}
\item A chirality with a diameter close to 1.2 nm, the maximal diameter of the diameter distribution.   
\item A metallic chirality without a gap at low energy.
\end{itemize}

Based on these arguments, we choose the chirality (9,9) as the second bound to compare with the experiment. The predictions for the second bound are shown as the blue dotted line in Fig.\ref{fig3}.

Fig.\ref{fig3}(a) and (b) show only a qualitative agreement with the surface conductivity model. Indeed, it underestimates the real part of the dielectric function in the range [0.4,4] eV and only the first bound matches with the experimental data for photon energy smaller than 0.35 eV.
On the opposite, the predictions from the \textit{ab initio} calculations, shown in Fig.\ref{fig3}(c) and (d), bracket the experimental data for the whole energy range explored in the experiment, except above 2.7~eV.
Since the SWCNT diameter distribution is relatively large, the individual peaks in the interband transitions originating from the band gaps, are strongly broadened and attenuated in the experiment.
In consequence, it is not possible to measure the band-gap values from the experimental data and compare them to the theoretical predictions.

We define a single quantifier to discriminate between the two theories: the screened plasma frequency $E^\star$ that can also be understood as a cut-off energy.
Indeed, for photon energy below this cut-off frequency, the m-SWCNT layer behaves as a metal. It has a dielectric function with a negative real-part.
On the opposite, above the cut-off energy, the real part of the dielectric function is positive. Consequently, to avoid the confusion with the plasma frequency shown in Fig.(\ref{fig1}-c) we will denote this quantity as the "cut-off energy".
It is defined as the frequency for which the real part of the dielectric function is null. It can be extracted from the experiment and compared to the theories as shown by the purple star and the grey line in Fig.(\ref{fig3}-a)-Fig.(\ref{fig3}-c). Based on Drude's model, we show in supplementary material S8 the relationship between the cut-off energy $E^\star$, the plasma frequency $\omega_p$ and the contribution of the interband transition in the axial $\ve_b^\pr$ and the transverse polarization $\ve_b^\bot$ evaluated at the cut-off energy. It reads: 
\begin{align}
	E^\star =\hbar \sqrt{\frac{\omega_p^2}{\ve_b^\pr+2\ve_b^\bot}-\gamma^2}
	\label{eq:cutoff}
\end{align}

Except for the loss rate $\gamma$ that is considered to be the same in the two theories, all other quantities differ as unambiguously shown by the theoretical study and highlighted in Fig.(\ref{fig1}-c) and Fig.(\ref{fig1}-d). Based on Eq.(\ref{eq:cutoff}) the cut-off energy can be considered as an excellent quantifier to discriminate the two theories.

\begin{figure}
	\includegraphics[width=8cm]{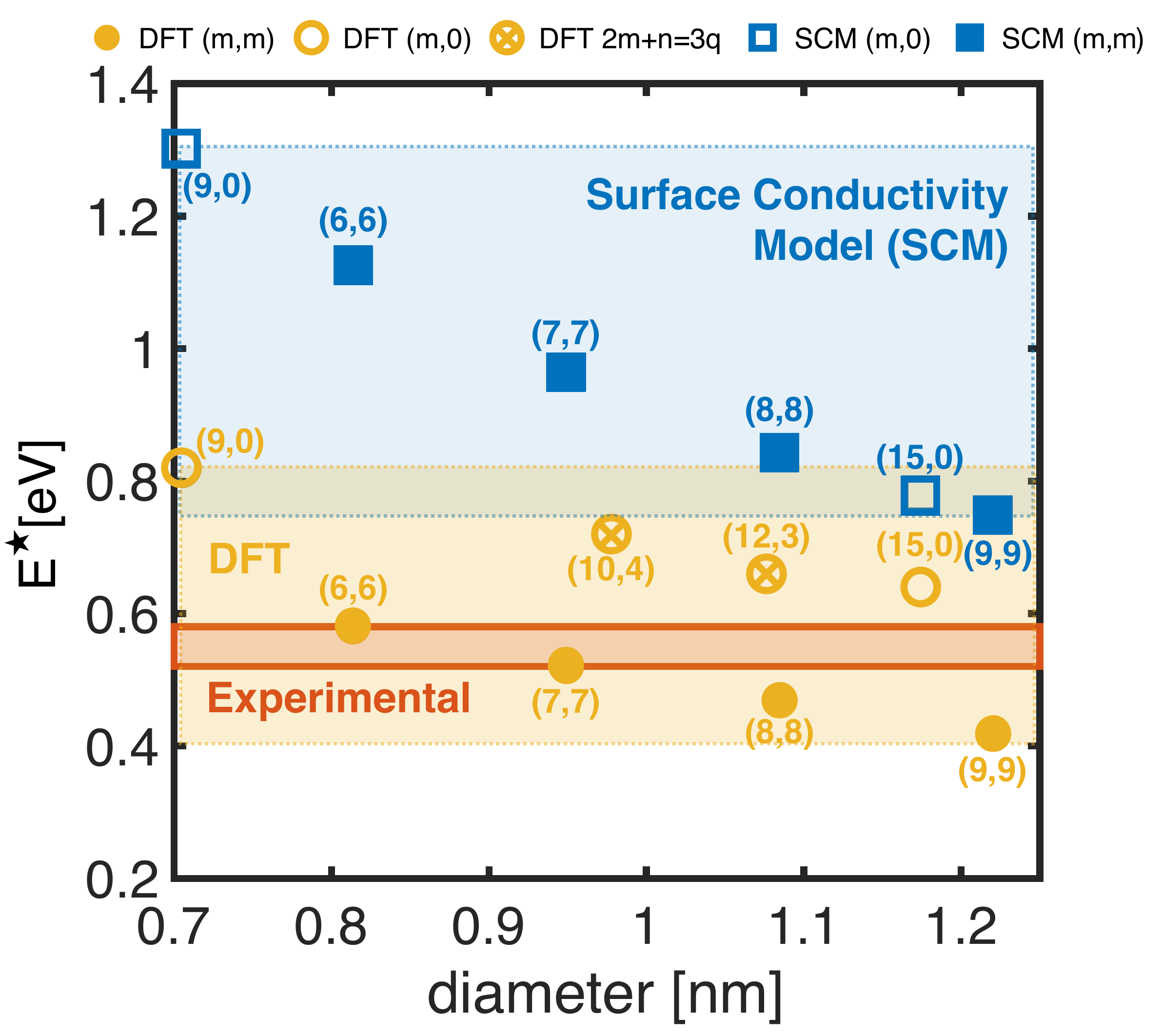}
	\caption{Cut-off energy $E^\star$ as a function of the m-SWCNT diameter. Experimental data are the red shaded area. Circles are predictions from the DFT calculations. They are delimited by the light-yellow shaded area. Squares are predictions from the linear surface conductivity model. They are delimited by the light-blue shaded area. Armchair (m,m), zigzag (m,0) and chiral (2m+n=3q) species are specified by the filled, open and crossed symbols, respectively. Chiral species could not be predicted by the linear surface conductivity model.}
	\label{fig4}
\end{figure}

The experimental value of the cut-off energy is $E^\star = 0.55 \pm 0.03$ eV. It is compared with the predictions given by the linear conductivity model and the DFT calculations in Fig.\ref{fig4}.
This figure displays $E^\star$ as a function of m-SWCNT diameter.
Experimental results are represented as a red shaded area, with a confidence interval of $\pm$~0.03~eV.
DFT calculations for various chiralities inside the experimental diameter distribution are represented by yellow circles, while values predicted by the linear surface conductivity model are represented by blue squares.
The upper bound is given by the chirality (9,9) in the two theories, while the lower bound is given by the chirality (9,0).

The experimental value is clearly within the interval defined by the DFT calculations and clearly outside the values predicted by the surface conductivity model, its lowest value being 0.17~eV too high.
It is very close to the predictions for chiralities (6,6) and (7,7) made by the DFT calculations, which more or less fall in the middle of the experimental diameter distribution.

To summarize, the cut-off energy as a quantifier proves unambiguously that the predictions from DFT calculations are more likely in agreement with the experiment, the linear conductivity surface model over-estimating the cut-off energy by 5 standard deviations. This discrepancy can be understood based on Eq.(\ref{eq:cutoff}).
As shown in Fig.(\ref{fig1}-c), DFT calculations predict smaller values of the plasma frequency than the surface conductivity model in the whole diameter range probed in this study. The difference in the plasma frequency predictions is a consequence of the relaxing of the set of assumptions (A2) performed by the linear surface conductivity model. The curvature radius of carbon nanotube not only leads to band gap at low energy but also has an effect on the value of the plasma frequency $\omega_p$. Furthermore, according to Eq.(\ref{eq:cutoff}), the quantity $\ve_b^\bot$ also contributes to a decrease of the cut-off energy as compared to the plasma frequency due to screening.
Following Fig.(\ref{fig1}-d), the decrease is expected to be more important for the DFT calculations than for the surface conductivity model since this model assumes that carbon nanotubes are non-polarisable objects in the plane transverse to the nanotube axis. This result highlights that the approximation (A1) is too crude to describe the optical properties of carbon nanotubes. Last but not least, DFT calculations take into account the curvature-induced low-energy gap existing in non-armchair SWCNT while the surface conductivity model does not.  As observed in Fig.\ref{fig4}, it leads to higher $E^\star$ for zigzag and chiral compared to armchair nanotubes, independently of the diameter effect. Consequently, it could not explain the discrepancy between the measurement and the predictions of the surface conductivity model which is mainly due to the failure of the approximation (A1) and the magnitude of the plasma frequency.

\section{Conclusion}
We reported the measurement of the complex dielectric constant of metallic carbon nanotubes films over a wide range of energies ( 0.05 to 4~eV) and the comparison with predictions from two theoretical models.
This allowed us to investigate intraband transitions in m-SWCNT and in particular to extract the cut-off frequency $E^\star = 0.55 \pm 0.03$ eV, below which a m-SWCNT film behaves like a metal. Based on this quantity, we highlight a correct agreement with DFT calculations and exclude prediction based on the analytical model. A similar conclusion was reached for graphene\cite{Cheon_2019}.
The assumptions performed in the analytical model\cite{Slepyan_1999} are too crude to describe correctly the optical properties of m-SWCNT.
Our methodology strengthens the need for \textit{ab initio} calculations to predict accurately the electronic and optical properties of SWCNT, which could have consequences for several applications in physics\cite{Lakhtakia_2006, Nemilentsau_2007,Rubi_2021} or biology\cite{Rubi_2022}.

\begin{acknowledgement}
ER and NI contributed equally to this work. They initiated and lead this research.
The samples were sorted by DB under the supervision of NI with insights from SC.
The reflectance measurements were performed by DB under the supervision of NI and JLB.
The Kramers-Kronig analysis were performed by DB and ER.
The DFT calculations were performed by PH.
ER and NI wrote a draft of the paper.
All authors discussed the results, agreed with the conclusions, and contributed to the final version of the letter.
Authors thanks M. Ramonda from CTM Montpellier for his help with AFM measurements.
Reflectance measurements were performed on the technological platform IRRAMAN of the University of Montpellier and authors thanks D. Maurin for his help with IR measurements.
\end{acknowledgement}

\begin{suppinfo}
Carbon nanotube Raman spectroscopy; UV-Visible absorption spectroscopy of sorted nanotube suspensions; extraction of nanotube complex refractive index from reflectance measurements; complex refractive index of the m-SWNT layer; transmittance measurement; comparison with a Drude model; DFT calculation details; relationship between cut-off energy and plasma frequency
\end{suppinfo}

\bibliography{acs-CNT_nano}

\end{document}